\documentclass[12pt]{article}
\usepackage[utf8]{inputenc}
\usepackage{mathtools}
\usepackage{amsmath}
\usepackage{amssymb}
\usepackage{mathptmx}
\usepackage{bm}
\usepackage[semicolon,sort&compress]{natbib}
\usepackage{geometry}
 \usepackage{authblk}
 \usepackage{titling}
 \usepackage[dvipsnames]{xcolor}

\definecolor{myred1}{rgb}{0.6, 0.01, 0.1}
\definecolor{myred2}{RGB}{219, 48, 122}
\definecolor{mypink3}{cmyk}{0, 0.7808, 0.4429, 0.1412}
\definecolor{mygray}{gray}{0.6}
\colorlet{myred3}{red!70!black}
\colorlet{myblue1}{blue!60!black}
\colorlet{myred4}{black!60!brown}
\usepackage{hyperref}
\hypersetup{colorlinks=true, linkcolor=myred1, citecolor = myblue1, urlcolor = myred4}
\usepackage{url}
\DeclareMathOperator{\Tr}{Tr} 
 
\title{Comments on: A new additive decomposition of velocity gradient, by B. Sun [Phys. Fluids 31, 061702 (2019)]}
\author{Abhijit Mitra}
\affil[]{\small{Department of Aerospace Engineering, Indian Institute of Science, Bangalore, India}}
\date{June 19, 2020}

\begin{document}
\maketitle

\begin{abstract}
    Comments on “A new additive decomposition of velocity gradient [Phys. Fluids 31, 061702 (2019)]" is presented.
\end{abstract}

\hrule
\vspace{5pt}

 The Cauchy-Stokes decomposition of the velocity gradient tensor into a symmetric strain-rate tensor $\mathbf{D}$ and an anti-symmetric spin tensor $\mathbf{W}$ is well-known \citep{Aris, Lumley, Kundu}. 
\begin{equation} \label{eq:irreduc}
 \mathbf{\nabla} \mathbf{u} = \mathbf{D} + \mathbf{W}
\end{equation}
The spin tensor $\mathbf{W}$ is the cartesian tensor representation of vorticity, $2\bm{\omega}$, in the three dimensional physical vector space, where $\mathbf{W}_{ij} = -\epsilon_{ijk}\omega_k$ ($\epsilon_{ijk}$ is the permutation tensor). \cite{Coope65} and \citet{Coope70} noted that a general second-rank tensor, like $\mathbf{\nabla u}$, can be decomposed into three unique, \textit{irreducible} second-rank tensors of various weights,
\begin{equation}\label{eq:irreduc1}
\mathbf{\nabla} \mathbf{u} = \mathbf{D_0} + \mathbf{U} +\mathbf{W} 
\end{equation}
 where $\mathbf{D_0}$ is a symmetric, traceless (called \textit{natural}) second-rank tensor of weight two. $\mathbf{W}$ is an anti-symmetric, second-rank tensor of weight one, whereas $\mathbf{U} \big(= \frac{1}{3} \Tr \big[\mathbf{\nabla u}\big]\mathbf{\delta} \big)$ is the zeroth-weight, second-rank isotropic tensor, with $\mathbf{\delta}$ being the second-rank unit tensor (the Kronecker delta). $\Tr \big[\mathbf{\nabla u}\big]$ is the trace (a scalar) of the velocity gradient tensor. $\mathbf{D_0,\,W}$ and $\mathbf{U}$ are all irreducible second-rank tensors under the three-dimensional rotational group $SO(3)$. The weight of an \textit{irreducible} tensor is $k$ if its dimension is $2k+1$. Dimensions of $\mathbf{D_0,\,W}$ and $\mathbf{U}$ are $5,\,3$ and $1$ respectively when represented by, say, an orthonormal basis in a real, three dimensional vector space (roughly speaking, the number of independent components of an irreducible second-rank tensor, in a real three-dimensional vector space, is its dimension). Any other possible decompositions are necessarily reducible. Undoubtedly, eq.(\ref{eq:irreduc1}) was known before the works of \cite{Coope65} and \citet{Coope70}, but their significance lie in the fact that they provide a general algorithm to find such irreducible decompositions of any arbitrary ranked cartesian tensor in three dimensional vector space, under the three-dimensional group of rotations $SO(3)$. \vspace{2mm}
 
\noindent One such reason of interest in decompositions of $\mathbf{\nabla u}$ is the need to identify vortices in fluid flows. In the quest to find characteristics that define a vortex, the  vorticity field has been found to be lacking due to a variety of reasons \citep{Epps2017}. An  interesting, alternative  proposition of a novel decomposition of the velocity gradient tensor is presented by \citet{Sun19} based on the Lie algebra of the special orthonormal Lie group $SO(3)$. This decomposes the velocity gradient tensor into a component which is a rotation tensor $\mathbf{Q}$ instead of the usual spin tensor. As a sidenote, \citet{Kundu, Lumley}, amongst many others, could be potential sources of confusion as $\mathbf{W}$ is called the rotation tensor in these references, whereas here it is called the spin tensor and the rotation tensor $\mathbf{Q}$ is an element of $SO(3)$ $\big( and\,\,\mathbf{W} \notin SO(3)\big)$.  \citet{Sun19} had noted that a deeper significance of this decomposition is not yet clear and further investigations are necessary in that direction. These comments are intended to interpret and rectify some aspects of \citet{Sun19}.\vspace{2mm}

\noindent \citet{Sun19} decomposes the velocity gradient tensor as, \begin{equation} \label{eq: Sun_decomp}
    \mathbf{\nabla} \mathbf{u} = \mathbf{K} + \mathbf{Q}
\end{equation}
where $\mathbf{Q} \in SO(3)$ is a rotation tensor  and $\mathbf{K}$ is the residual. It has to be noted that this decomposition is not irreducible under $SO(3)$. Anti-symmetric tensors like $\mathbf{W}$ belong to the Lie algebra $so(3)$ of the Lie group $SO(3)$. There exists an exponential map from $so(3) \to SO(3)$. Exploiting this, \citet{Sun19}  expresses a rotation tensor $\mathbf{Q} \in SO(3)$ as, 

 \begin{equation} \label{eq:expo_map}
     \mathbf{Q} = e^{\mathbf{W}}
 \end{equation}
First, \citet{Sun19} does not address the issue of dimensional inconsistency in eq.(\ref{eq:expo_map}). Physical dimension of $\mathbf{W}$ is sec$^{-1}$ - there are obvious problems and one cannot exponentiate a dimensional quantity. It is unclear if all the physical quantities in \citet{Sun19} are non-dimensional. Second, presuming that $\mathbf{\nabla u},\, \mathbf{D}$ and $\mathbf{W}$ are non-dimensional right from the outset, $\mathbf{K}$ and $\mathbf{Q}$ (in decomposition(\ref{eq: Sun_decomp})) are not irreducible and can still be further reduced, ultimately leading to decomposition(\ref{eq:irreduc1}). Third, there exists a more pressing problem with eq.(\ref{eq:expo_map}). $\mathbf{Q}$ does not represent a \textit{one-parameter subgroup} of $SO(3)$ in the neighborhood of $\mathbf{I}$ \citep{Fegan, Hall}, where $\mathbf{I}$ is identity element of $SO(3)$. This is an essential requirement for an isomorphism from $so(3)$ to $SO(3)$ in the neighborhood of identity. The third objection is fundamental because fixing it will seamlessly fix the first objection and not vice-versa. In other words, even if all the quantities were dimensionless, eq.(\ref{eq:expo_map}) would still not represent a one-parameter subgroup of $SO(3)$. Equation(\ref{eq:expo_map}) needs rectification. Otherwise, it would result in discrepancies in the inferences that can be drawn from Sun's exposition. For example, a question raised by \citet{Sun19}: under what condition(s)  $\mathbf{K}$ is symmetric? To further his arguments, $\mathbf{K}$ can be symmetric, at best, for vortical flows with vanishing vorticity ($\omega \to 0$). It will, however, be shown here that it is impossible for $\mathbf{K}$ to be symmetric in a flow with vorticity, if $\mathbf{Q}$ is represented correctly as a one-parameter subgroup of $SO(3)$. Such inconsistencies occur due to the disregard of this fundamental property of the mapping from Lie algebras to their corresponding Lie groups. A natural justification for the consideration of a one-parameter subgroups is provided in what follows along with the implications and limitations of this decomposition. \vspace{2mm}
 
  A Lie group, such as the $SO(3)$, has the structure of a differentiable manifold in the vector space of real matrices. On any integral curve induced by the tangent tensor field like $\mathbf{W}$ on $SO(3)$, the following holds \citep{Hall} in the neighborhood of $\mathbf{I}$,
  \begin{equation}\label{eq:Dyn_eq}
      \frac{d\sigma(\tau)}{d\tau} = \mathbf{W}
  \end{equation}
  where $\tau$ is the parameter in the map $\sigma : I_R \to SO(3)$, with $\tau \in I_R = [a,b] \in \mathbb{R}$ and $I_R$ contains 0 ($a,b \in \mathbb{R}$). $\tau$ might be interpreted as the time increment/decrement, $t-t_0$, where $\sigma(0) = \mathbf{I}$ for some reference time $t_0$. Along the integral curve $\sigma(\tau) \in SO(3)$, eq.(\ref{eq:Dyn_eq}) demands,
  \begin{equation}\label{eq:Q_exp}
     \sigma(\tau)= \mathbf{Q}(\tau) = e^{\mathbf{W}\tau}
 \end{equation} 
   Equation(\ref{eq:Q_exp}) would describe a family of rotations parameterised by $\tau$: \textit{a one-parameter subgroup of $SO(3)$}. Equation(\ref{eq:Q_exp}) can also be derived by a much simpler consideration of the orthonormal, rotation tensor $\mathbf{Q}(\tau)$. Time derivative of $\mathbf{Q}\mathbf{Q}^T$ ($=\mathbf{I}$) is 
   \begin{equation} \label{eq:q_qdot}
       \frac{d(\mathbf{Q}\mathbf{Q}^T)}{d\tau} = \dot{\mathbf{Q}}\mathbf{Q}^T + \mathbf{Q}\dot{\mathbf{Q}}^T = \dot{\mathbf{I}} = \mathbf{0}
   \end{equation}
   where $\dot{\mathbf{Q}}, \dot{\mathbf{I}}$ denote the time derivatives of $\mathbf{Q}$ and $\mathbf{I}$ respectively, and $\mathbf{Q}^T$ is the transpose of $\mathbf{Q}$ (with $\mathbf{Q}^T =\mathbf{Q}^{-1}$). From eq.(\ref{eq:q_qdot}), it is obvious that $\dot{\mathbf{Q}}\mathbf{Q}^T$ is anti-symmetric. Thus, for any $\mathbf{Q}$ there always exists a  $\dot{\mathbf{Q}}$ such that,
   \begin{equation}\label{eq:Q_dot}
       \dot{\mathbf{Q}} = \mathbf{W Q}
   \end{equation}
   This tensorial differential equation is equivalent to eq.(\ref{eq:Dyn_eq}), and the following satifies eq.(\ref{eq:Q_dot}),
   \begin{equation}\label{eq: correct_expo}
       \mathbf{Q}(\tau) = e^{\mathbf{W}\tau}\mathbf{Q}_0
   \end{equation}
   Consider the integral curve in $SO(3)$ through the identity with  $\mathbf{Q}_0 = \mathbf{I}$, thereby reducing eq.(\ref{eq: correct_expo}) to eq.(\ref{eq:Q_exp}), reiterating the fact that $\mathbf{Q}(\tau)$ is a one-parameter sub-group of $SO(3)$ near $\mathbf{I}$. This is mathematically and dimensionally a more consistent and correct exponential map from $so(3) \to SO(3)$ than eq.(\ref{eq:expo_map}). If $\mathbf{W}$ is independent of time, there are no restrictions on $\tau$ in eq.(\ref{eq:Q_exp}), and  eq.(\ref{eq:expo_map}) is recovered only for a special case of $\tau = 1$. But, in a generic fluid flow field, $\mathbf{W}$ must be a function of time for a material fluid parcel. Therefore, this limits the validity of eq.(\ref{eq:Q_exp})  to $|\tau | \to 0$, wherein the allowable limit of $|\tau |$ is much smaller than the time-scale of any appreciable change in $\mathbf{W}$. \vspace{2mm}
   
   \noindent  The Rodrigues' formula used by \citet{Sun19} (eq.16 of the paper) is valid only for $\tau = 1$. Based on this modified exponential map in eq.(\ref{eq:Q_exp}), the complete Rodrigues' formula for $\mathbf{Q}$ is,
   $$\mathbf{Q} = \mathbf{I} + \frac{\sin \omega \tau}{\omega}\mathbf{W} + \frac{1-\cos\omega \tau}{\omega^2}\mathbf{W}^2 $$
   And, if decomposition (\ref{eq: Sun_decomp}) for a non-dimensional $\mathbf{\nabla u}$ is demanded such that $\mathbf{K}$ is symmetric, then the following must hold,
   \begin{equation}\label{eq:K_sym_1}
       \big(1-\frac{\sin \omega \tau}{\omega}\big) \mathbf{W} = \bm{0}
   \end{equation}
  Equation(\ref{eq:K_sym_1}) can be satisfied for any allowable $\tau$ ($|\tau | \to 0$), if and only if $\mathbf{W} =\mathbf{0}$ identically. Thus, $\mathbf{K}$ can never be symmetric in a vortical flow. This is in distinction to the possibility of a symmetric $\mathbf{K}$ from Sun's exposition, where symmetric $\mathbf{K} $ is allowable for vortical flows with $\omega \to 0$. It is clear that incorrect use of the transformation from the Lie algebra to Lie group, $so(3) \to SO(3)$ (eq.(\ref{eq:expo_map})) is the source of such discrepancies. \vspace{2mm}
  
  \noindent For a turbulent flow, as mentioned earlier, $\mathbf{W}$ would have erratic dependence on time, and the exponential map would be valid just for infinitesimal time durations, i.e., $|\tau| \to 0$. In that limit, $\mathbf{Q} = \mathbf{I}$ and $\mathbf{K} = \mathbf{D}+\mathbf{W} -\mathbf{I}$, severely restricting the applicability of this new and not irreducible decomposition of the velocity gradient tensor.  
\bibliographystyle{unsrtnat}
\bibliography{my_references}
\end{document}